\documentclass[aps, prb, reprint, superscriptaddress]{revtex4-1}
\usepackage{amsmath}
\usepackage{amsfonts}
\usepackage{graphicx}
\usepackage{natbib}
\usepackage{amssymb} %for circular arrows
\usepackage{color}

\begin{document}

\title{Evaluating arbitrary strain configurations and doping in graphene with Raman spectroscopy}

\author{Niclas S. Mueller}
\email{niclasm@physik.fu-berlin.de}
\affiliation{Freie Universit\"at Berlin, Department of Physics, Arnimallee 14, D-14195 Berlin, Germany}
\author{Sebastian Heeg}
\affiliation{School of Materials, The University of Manchester, Manchester M13 9PL, UK}
\affiliation{Photonics Laboratory, ETH Z\"{u}rich, 8093 Z\"{u}rich, Switzerland}
\author{Miriam Pe\~{n}a Alvarez}
\affiliation{J. Heyrovsk\'{y} Institute of Physical Chemistry, Academy of Sciences of the Czech Republic, Dolej\v{s}kova 3, CZ-18223 Prague 8, Czech Republic}
\author{Patryk Kusch}
\affiliation{Freie Universit\"at Berlin, Department of Physics, Arnimallee 14, D-14195 Berlin, Germany}
\author{S\"oren Wasserroth}
\affiliation{Freie Universit\"at Berlin, Department of Physics, Arnimallee 14, D-14195 Berlin, Germany}
\author{Nick Clark}
\affiliation{School of Materials, The University of Manchester, Manchester M13 9PL, UK}
\author{Fred Schedin}
\affiliation{Centre for Mesoscience and Nanotechnology, The University of Manchester, Manchester M13 9PL, UK}
\author{John Parthenios}
\affiliation{Institute of Chemical Engineering Sciences, Foundation of Research and Technology - Hellas, Patras 26504, Greece}
\author{Konstantinos Papagelis}
\affiliation{Institute of Chemical Engineering Sciences, Foundation of Research and Technology - Hellas, Patras 26504, Greece}
\affiliation{Department of Physics, University of Patras, Patras 26504, Greece}
\author{Costas Galiotis}
\affiliation{Institute of Chemical Engineering Sciences, Foundation of Research and Technology - Hellas, Patras 26504, Greece}
\affiliation{Department of Chemical Engineering, University of Patras, Patras 26504, Greece}
\author{Martin Kalb\'{a}\v{c}}
\affiliation{J. Heyrovsk\'{y} Institute of Physical Chemistry, Academy of Sciences of the Czech Republic, Dolej\v{s}kova 3, CZ-18223 Prague 8, Czech Republic}
\author{Aravind Vijayaraghavan}
\affiliation{School of Materials, The University of Manchester, Manchester M13 9PL, UK}
\affiliation{National Graphene Center, The University of Manchester, Manchester M13 9PL, UK}
\author{Uwe Huebner}
\affiliation{Leibnitz Institute of Photonic Technology, 07745 Jena, Germany}
\author{Roman Gorbachev}
\affiliation{School of Physics \& Astronomy, The University of Manchester, Manchester M13 9PL, UK}
\author{Otakar Frank}
\affiliation{J. Heyrovsk\'{y} Institute of Physical Chemistry, Academy of Sciences of the Czech Republic, Dolej\v{s}kova 3, CZ-18223 Prague 8, Czech Republic}
\author{Stephanie Reich}
\email{reich@physik.fu-berlin.de}
\affiliation{Freie Universit\"at Berlin, Department of Physics, Arnimallee 14, D-14195 Berlin, Germany}

\date{\today}

\begin{abstract}
Raman spectroscopy is a powerful tool for characterizing the local properties of graphene. Here, we introduce a method for evaluating unknown strain configurations and simultaneous doping. It relies on separating the effects of hydrostatic strain (peak shift) and shear strain (peak splitting) on the Raman spectrum of graphene. The peak shifts from hydrostatic strain and doping are separated with a correlation analysis of the 2D and G frequencies. This enables us to obtain the local hydrostatic strain, shear strain and doping without any assumption on the strain configuration prior to the analysis. 
We demonstrate our approach for two model cases: Graphene under uniaxial stress on a PMMA substrate and graphene suspended on nanostructures that induce an unknown strain configuration. We measured $\omega_\mathrm{2D}/\omega_\mathrm{G} = 2.21 \pm 0.05$ for pure hydrostatic strain. Raman scattering with circular corotating polarization is ideal for analyzing strain and doping, especially for weak strain when the peak splitting by shear strain cannot be resolved. 
\end{abstract}

\maketitle

\newpage
\section{Introduction}
Being the first two-dimensional material discovered, graphene has attracted a lot of attention for its extraordinary properties\cite{Geim2007}. Ultra-high carrier mobility, large thermal and electrical conductivity, impermeability to any gases and extreme mechanical robustness are combined in one material \cite{Novoselov2012}. Two-dimensional materials like graphene are strongly affected by their environment and distortions introduced during their processing, which affects their performance in devices. Locally varying strain is induced in the graphene lattice upon deposition on a substrate \cite{Ishigami2007, Couto2014, Neumann2015}, thermal annealing \cite{Bao2009} and stacking with other 2D materials \cite{Ahn2013, Woods2014}. The substrate and adsorbed or intercalated substances typically induce a Fermi level shift and dope graphene \cite{Ryu2010, Kim2013}. Strain and doping both affect the electronic and chemical properties of graphene. 

Raman spectroscopy is a powerful and non-destructive tool for characterizing the local properties of graphene \cite{Ferrari2006,Malard2009}. Spectral position, width and intensity of the Raman modes give information about layer number, defects, doping and strain in the graphene lattice \cite{Hao2010, Lucchese2010, Das2008, Mohiuddin2009, Huang2009}. However, these characteristics usually vary simultaneously, which hinders a straightforward analysis of the Raman spectra \cite{Bruna2014, Lee2012, Froehlicher2015}. Doping and strain both lead to a shift of the Raman modes and strain additionally to a peak splitting \cite{Das2008, Mohiuddin2009, Huang2009}. It was suggested to separate the contributions of strain and doping with a correlation analysis of the G and 2D frequencies \cite{Lee2012}. This approach has been extensively used for accessing the local properties of graphene on various substrates \cite{Ahn2013, Lee2014, Frank2014, Droegeler2014, McCreary2014, Neumann2015, Anagnostopoulos2015, Bronsgeest2015, Godel2017}.

\begin{figure}
\centerline{\includegraphics{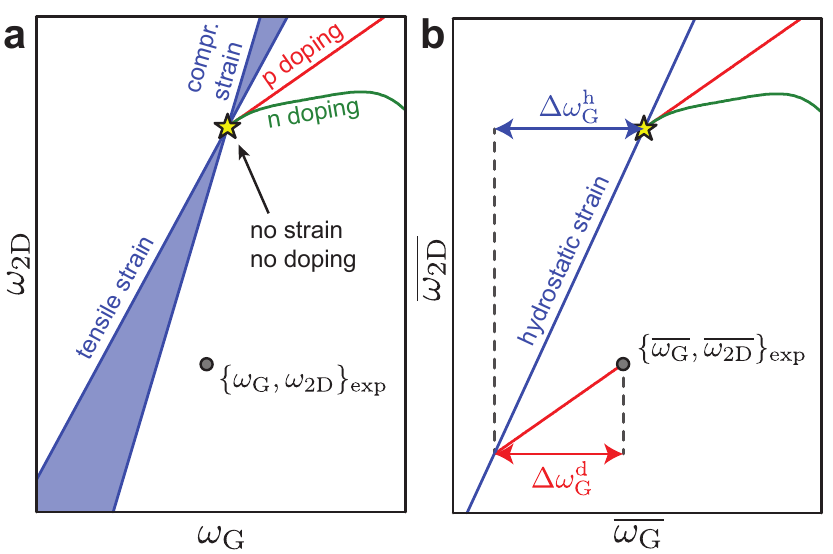}}
\caption{Correlation analysis of 2D and G frequencies to separate frequency shifts from strain and doping. (a) Reference values for strain shift rates deviate from each other and depend on the underlying strain configuration. The frequency shifts from strain and doping cannot be separated for a potential data point $\lbrace \omega_\mathrm{G}, \omega_\mathrm{2D} \rbrace_\mathrm{exp}$. (b) When calculating the mean frequencies $\overline{\omega_\mathrm{G}}$ and $\overline{\omega_\mathrm{2D}}$, the frequency shifts expected for pure hydrostatic strain can be used as reference values. The separation of G frequency shifts from doping $\Delta \omega_\mathrm{G}^\mathrm{d}$ and strain $\Delta \omega_\mathrm{G}^\mathrm{h}$ is exemplarily shown for the case of p doping.}
\vspace{-1mm}
\label{fig:IntroFigure}
\end{figure}

A sketch of the correlation analysis is shown in Fig.\ \ref{fig:IntroFigure}a. The frequencies expected for purely strained and purely doped graphene are plotted as lines resembling a coordinate system in the $\omega_\mathrm{2D}$-$\omega_\mathrm{G}$ plot. Extracting strain and doping for a potential frequency pair $\lbrace \omega_\mathrm{G},\omega_\mathrm{2D} \rbrace_\mathrm{exp}$ is done by projecting the point onto the strain and doping axes. The slope $\omega_\mathrm{2D}/\omega_\mathrm{G}$ for pure doping is $< 1$ because the G frequency depends much stronger on doping than the 2D frequency \cite{Das2008, Das2009, Bruna2014, Froehlicher2015}. This is mostly attributed to a nonadiabatic Kohn anomaly at the $\Gamma$ point in the phonon dispersion relation \cite{Lazzeri2006}. Strain leads to a slope of $\sim 2$ since the 2D mode is a phonon overtone that shifts with approximately twice the rate of a single phonon excitation (G line). The exact value of the strain-related slope, however, depends very sensitively on the strain configuration. Values ranging from 2-3 were reported for uniaxial and biaxial strain; they will scatter even more for other strain configurations \cite{Mohiuddin2009, Yoon2011, Ding2010, Berciaud2014, Metzger2010, Zabel2012, Mohr2010}. As one is typically unaware of the underlying strain configuration, separating the frequency shifts from strain and doping becomes impossible (see e.g.\ the $\omega_\mathrm{2D}$, $\omega_\mathrm{G}$ pair in Fig.\ \ref{fig:IntroFigure}a). A Raman scattering based approach for strain-doping analysis irrespective of the strain configuration is highly desirable.

Here, we show how to extract strain and doping from the Raman spectra for any (unknown) strain configuration and simultaneous doping. The peak splitting from shear strain is removed from the spectra by calculating the mean frequencies of the G$^-$ and G$^+$ as well as the 2D$^-$ and 2D$^+$ modes. The frequency shifts from hydrostatic strain and doping are separated with a correlation analysis of the G and 2D mean frequencies. The strain and doping levels are calculated from the G frequency shifts compared to pristine graphene. We demonstrate two experimental approaches of separating the peak shifts induced by hydrostatic strain and doping and the splitting induced by shear strain. Recording Raman spectra with circularly polarized light is ideal for extracting the hydrostatic strain component at small strain where no peak splitting is visible. This approach is demonstrated for graphene under uniaxial stress. 
We then apply our strain analysis to a complex case of graphene covering nanostructures for which the strain configuration is totally unknown. Graphene is suspended on lithographically fabricated gold nanodiscs forming dimers, which induce strong local and non-uniform strain in the graphene bridging the gap between the two nanoparticles. Surface-enhanced Raman scattering probes the local strain and doping levels on the nanoscale. The two examples serve as model cases for unraveling the strain and doping levels of graphene on arbitrary substrates.

\section{Methodology}
Let us consider a general, unknown strain tensor 
\begin{equation}
\varepsilon(\mathbf{r}) = \left( \begin{array}{cc}
\varepsilon_{xx}( \mathbf{r}) & \varepsilon_{xy}( \mathbf{r}) \\ 
\varepsilon_{yx}( \mathbf{r}) & \varepsilon_{yy}( \mathbf{r})
\end{array} \right),
\end{equation}
with in-plane components $\varepsilon_{ij}$. By solving the secular equation, we calculate the phonon shift rates as \cite{Ganesan1970, Huang2009,Mohiuddin2009, Reich2000}
\begin{equation}
\begin{split}
\label{eq:PeakShiftRates}
\Delta \omega_\mathrm{pn}^\pm = &- \omega_\mathrm{pn}^0 \gamma_\mathrm{pn}\left( \varepsilon_{xx} + \varepsilon_{yy} \right) \\& \pm \frac{1}{2} \omega_\mathrm{pn}^0 \beta_\mathrm{pn} \sqrt{\left( \varepsilon_{xx} - \varepsilon_{yy} \right)^2 + 4 \varepsilon_{xy}^2},
\end{split}
\end{equation}
with the phonon frequency in the absence of strain $\omega_\mathrm{pn}^0$, the Grueneisen parameter $\gamma_\mathrm{pn}$ and the shear-strain phonon deformation potential $\beta_\mathrm{pn}$. Any strain configuration can be decomposed into a hydrostatic component $\varepsilon_\mathrm{h} = \varepsilon_{xx} + \varepsilon_{yy}$ and a shear component $\varepsilon_\mathrm{s} = \sqrt{\left( \varepsilon_{xx} - \varepsilon_{yy} \right)^2 + 4 \varepsilon_{xy}^2}$ (assuming $\varepsilon_{xy} = \varepsilon_{yx}$) \cite{Reich2000}. Hydrostatic strain corresponds to an isotropic increase or decrease in the size of the graphene lattice; it leads to a frequency shift $\Delta \omega_\mathrm{pn}^\mathrm{h} = -\omega_\mathrm{pn}^0 \gamma_\mathrm{pn} \varepsilon_\mathrm{h}$ determined by the Grueneisen parameter $\gamma_\mathrm{pn}$ \cite{Zabel2012}. Note that from its definition $\varepsilon_\mathrm{h} = \varepsilon_{xx} + \varepsilon_{yy}$ the hydrostatic strain is twice as large as the corresponding biaxial strain $\varepsilon_\mathrm{b} = \varepsilon_{xx} = \varepsilon_{yy}$. Shear strain corresponds to an anisotropic distortion of the graphene lattice leaving the area of the unit cell unchanged. It leads to a peak splitting $\Delta \omega_\mathrm{pn}^\mathrm{s} = \omega_\mathrm{pn}^0 \beta_\mathrm{pn} \varepsilon_\mathrm{s}$ that depends on the shear deformation potential $\beta_\mathrm{pn}$, while the mean position of the two peak components remains constant \cite{Thomsen2002, Mohiuddin2009}. Equation (\ref{eq:PeakShiftRates}) is strictly valid for first-order Raman processes such as the G mode. For higher-order processes like the 2D mode, the electronic structure and its dependence on strain have to be considered as well. The general concept of peak splitting by shear strain and peak shift by hydrostatic strain, however, also applies to the 2D mode, as was shown experimentally and theoretically \cite{Yoon2011, Huang2010, Zabel2012, Frank2011, Narula2012}. In the following, the peak components are labeled G$^-$ and G$^+$ for the G mode and 2D$^-$ and 2D$^+$ for the 2D mode.

The Raman response of graphene subjected to specific strain configurations, i.e.\ uniaxial and biaxial strain, has been extensively studied \cite{Mohiuddin2009,Huang2009,Huang2010, Yoon2011, Zabel2012, Narula2012, Corro2015, Frank2011, Soldatov2012}. For biaxial strain the diagonal tensor components fulfill $\varepsilon_{xx} = \varepsilon_{yy}$; for uniaxial strain they are connected by the Poisson ratio $\nu$ as $\varepsilon_{xx} = -\nu \varepsilon_{yy}$ (stress applied along $x$ axis) \cite{Mohiuddin2009}. The off-diagonal components vanish in both cases. The frequency shifts for these very specific strain configurations were previously used as reference values for the correlation analysis of G and 2D frequencies \cite{Lee2012, Ahn2013, Lee2014, Frank2014, Droegeler2014, McCreary2014, Neumann2015, Anagnostopoulos2015, Bronsgeest2015}. This implicitly assumed that either biaxial or uniaxial strain is present in the system under study. As the strain configuration is generally unknown and varies accross the sample, we suggest a different approach. We propose a methodology for the evaluation of arbitrary strain configurations and simultaneous doping in graphene:
\begin{enumerate}
\item The correlation analysis will be based on the phonon frequency shift induced by the hydrostatic strain component. To obtain it, we eliminate the peak splitting from shear strain $\Delta \omega_\mathrm{s}$ by calculating the mean frequencies
\begin{equation}
\overline{\omega_\mathrm{G}} = \frac{\omega_{\mathrm{G}^-}+\omega_{\mathrm{G}^+}}{2}, \quad \overline{\omega_\mathrm{2D}} = \frac{\omega_{\mathrm{2D}^-}+\omega_{\mathrm{2D}^+}}{2}.
\end{equation}
This leads to a data point in the correlation plot that is only affected by hydrostatic strain and doping (see Fig.\ \ref{fig:IntroFigure}b). 
\item The peak shifts from hydrostatic strain $\Delta \omega_\mathrm{pn}^\mathrm{h}$ and doping $\Delta \omega_\mathrm{pn}^\mathrm{d}$ are separated with a correlation analysis of $\overline{\omega_\mathrm{G}}$ and $\overline{\omega_\mathrm{2D}}$, following Ref.\ \citenum{Lee2012}. The slope expected for hydrostatic strain is used as a reference (illustrated in Fig.\ \ref{fig:IntroFigure}b).
\item Hydrostatic strain $\varepsilon_\mathrm{h}$ is calculated from the G mode shift $\Delta \omega_\mathrm{G}^\mathrm{h}$ and doping from $\Delta \omega_\mathrm{G}^\mathrm{d}$. 
\item Shear strain $\varepsilon_\mathrm{s}$ is calculated from the G splitting $\Delta \omega_\mathrm{G}^\mathrm{s}$.
\end{enumerate}
The approach requires no assumption on the strain configuration as an input parameter. However, the type of doping needs to be known for choosing the correct reference values. It can be obtained from reference measurements with other techniques; for many graphene-material combinations it may be found in literature. 

\section{Analysis with circular light polarization}
\begin{figure}
\centerline{\includegraphics{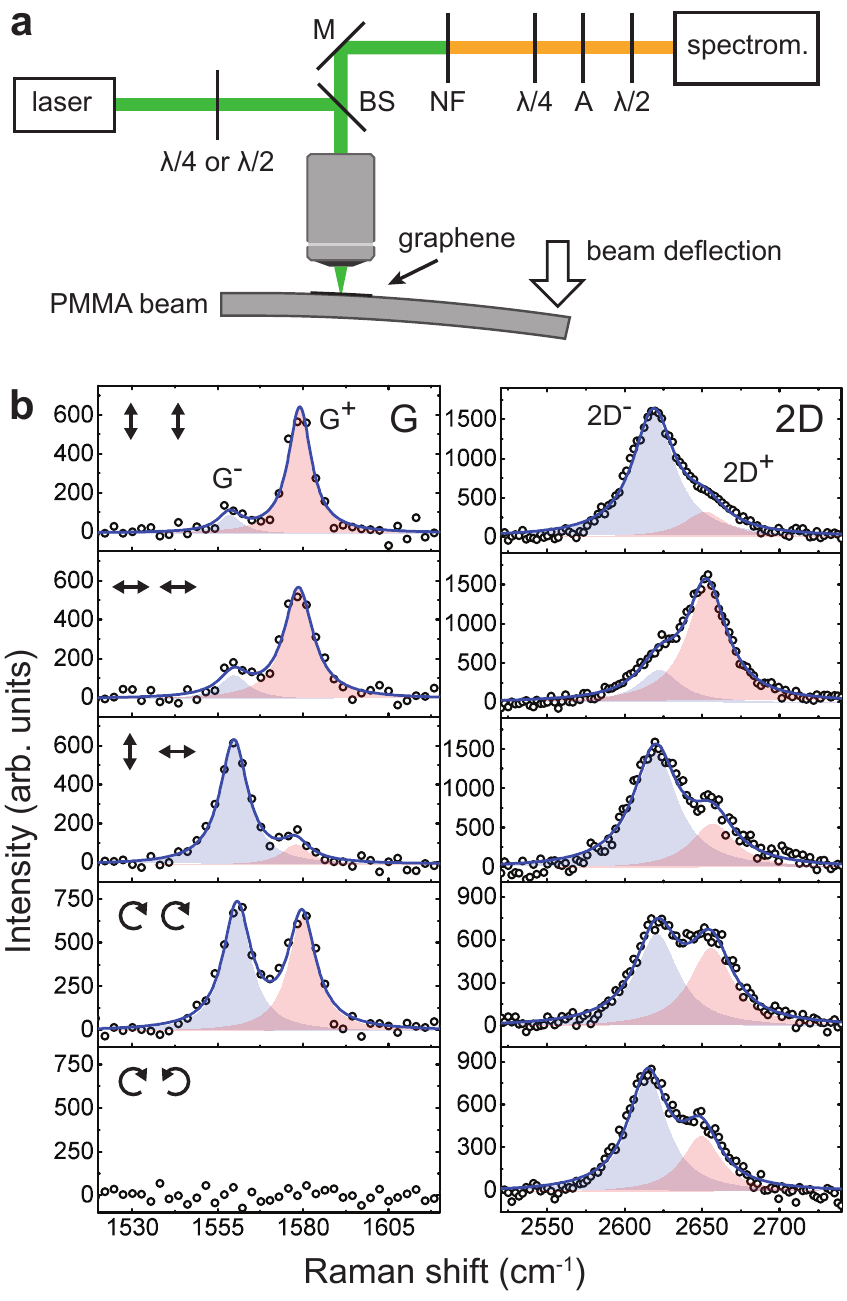}}
\caption{Polarized Raman spectroscopy of uniaxially-strained graphene. (a) Backscattering Raman setup with control of light polarization ($\lambda/4$ - quarter-wave plate, $\lambda/2$ - half-wave plate, BS - beam splitter, M - mirror, NF - notch filter, A - analyzer). For circular light polarization, two $\lambda/4$ plates, an analyzer and one $\lambda/2$ plate are used. (b) Raman spectra of the G and 2D mode of uniaxially-strained graphene (laser wavelength 532 nm). Fits with two Lorentzian peaks are superimposed on the experimental data. G$^-$ and 2D$^-$ peaks are colored blue, G$^+$ and 2D$^+$ peaks are colored red. Polarizations of the incoming- and outgoing light are indicated by arrows ($\updownarrow$ - linear along strain, $\leftrightarrow$ - linear perpendicular to strain, $\circlearrowright\circlearrowright$ - circular corotating, $\circlearrowright\circlearrowleft$ - circular contrarotating).}
\label{fig:Figure1}
\end{figure}

In the following, we present two examples for how to apply our methodology. These can be viewed as model cases for the strain analysis of graphene on arbitrary substrates. As a first example, we show that circular light polarization is ideal for measuring the mean frequencies $\overline{\omega_\mathrm{G}}$ and $\overline{\omega_\mathrm{2D}}$. We use a Raman setup where linear and circular light polarization can be independently chosen for incoming and scattered light (Fig.\ \ref{fig:Figure1}a, see Methods for details). Tensile uniaxial strain is induced in exfoliated monolayer graphene flakes by the deflection of a PMMA beam (Fig.\ \ref{fig:Figure1}a) \cite{Tsoukleri2009, Frank2010, Galiotis2015, Mohiuddin2009, Corro2015}. We observe a shift of the phonon frequencies to lower wavenumber because of hydrostatic strain and a peak splitting because of shear strain (Fig.\ \ref{fig:Figure1}b). For linearly polarized incoming and scattered light, the $+/-$ components differ strongly in intensity for different polarization configurations (Fig.\ \ref{fig:Figure1}b, $\updownarrow\updownarrow$, $\leftrightarrow\leftrightarrow$ and $\updownarrow\leftrightarrow$). From the intensity ratio of G$^-$ and G$^+$ peak, we find that the uniaxial strain direction is primarily along the zigzag direction with a misorientation of 7.5\% [see Methods, Eq.\ (\ref{eq:GmodeIntLinLightPol})]\cite{Mohiuddin2009,Huang2009}. 

The G$^-$ and $G^+$ peak are of equal intensity for circular corotating polarization as expected from the selection rules (Fig.\ \ref{fig:Figure1}b,  $\circlearrowright\circlearrowright$, see Methods for discussion). A similar behavior was observed for the 2D mode. Both G$^-$ and G$^+$ modes vanished for circular contrarotating polarization (Fig.\ \ref{fig:Figure1}b,  $\circlearrowright\circlearrowleft$). For this polarization, the 2D mode split into components of unequal intensity; the 2D$^-$ component was consistently more intense than the 2D$^+$ component. A similar behavior was visible for strain along the armchair direction in the graphene lattice (Supplementary Fig.\ S1).

\begin{figure}
\centerline{\includegraphics{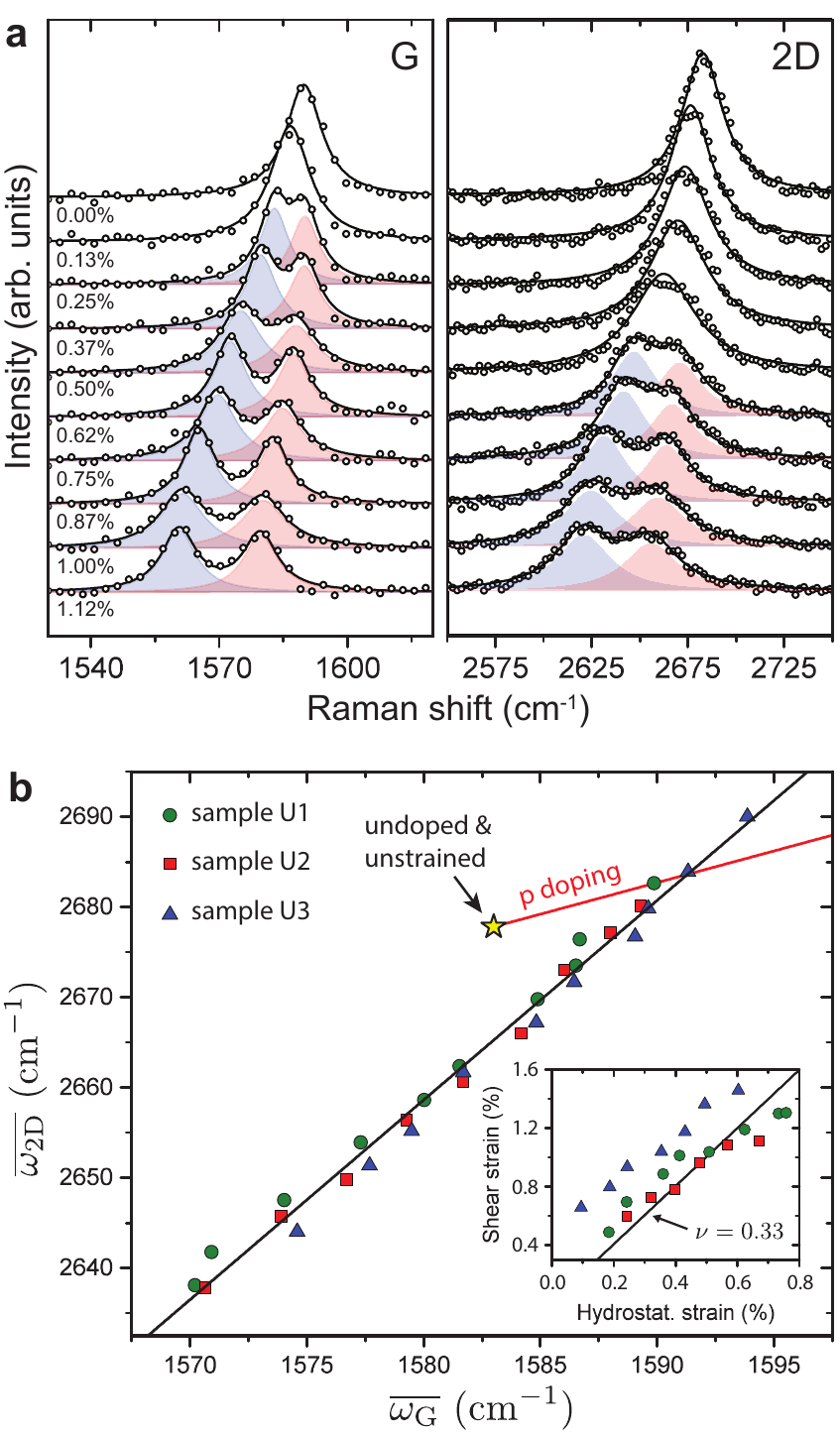}}
\caption{Extracting the hydrostatic strain, shear strain and doping from the Raman spectra of uniaxially-strained graphene using circular corotating polarization (laser wavelength 532 nm). (a) G and 2D mode of graphene under uniaxial strain which increases incrementally from top to bottom (sample U1). Fits with one or two Lorentzian peaks are superimposed on the data. G$^-$ and 2D$^-$ peaks are colored blue, G$^+$ and 2D$^+$ peaks are colored red. Plots are labeled by the expected uniaxial strain $\varepsilon_\mathrm{u}$ calculated from the beam deflection. (b) Correlation plot of 2D and G mean frequencies for three samples (U1-U3). A linear fit is plotted as a solid line. The red line indicates the expected spectral positions if only p-type doping is present (slope of 0.7, see Ref.\ \citenum{Lee2012} and data in Refs.\ \citenum{Das2008, Das2009}). The yellow star indicates spectral position in the absence of doping and strain (extracted from Ref.\ \citenum{Das2008}, 514 nm laser excitation, assuming a 2D mode dispersion of 100 cm$^{-1}$/eV). The inset shows shear strain as a function of hydrostatic strain. The expected slope for a Poisson ratio $\nu = 0.33$ is shown as a black line.}
\label{fig:Figure2}
\end{figure}

Recording Raman spectra with circular corotating light is ideal for measuring the mean frequencies $\overline{\omega_\mathrm{G}}$ and $\overline{\omega_\mathrm{2D}}$ because both peak components are equally intense. For linear light polarization, one of the peak components can vanish and the mean frequencies cannot be obtained. We demonstrate our approach of using circular light polarization for incrementally increasing uniaxial strain in Fig.\ \ref{fig:Figure2}a. The expected uniaxial strain $\varepsilon_\mathrm{u} = \varepsilon_\mathrm{h}/(1-\nu) = \varepsilon_\mathrm{s}/(1+\nu)$ is calculated from the beam deflection (see plot labels in Fig.\ \ref{fig:Figure2}a) \cite{Tsoukleri2009}. Both G and 2D modes were fit by two Lorentzian peaks of same spectral width. We thereby obtained the frequencies of all components, i.e.\ $\omega_{\mathrm{G}^-}$, $\omega_{\mathrm{G}^+}$, $\omega_{\mathrm{2D}^-}$ and $\omega_{\mathrm{2D}^+}$, which were used to calculate $\overline{\omega_\mathrm{G}}$ and $\overline{\omega_\mathrm{2D}}$. For low strain levels, when no peak splitting was visible, we obtained the mean frequencies from single Lorentzian peak fits. This is only possible because the strain-split G and 2D components have equal intensity, i.e.\ because we use circular-corotating light polarization.

In Fig.\ \ref{fig:Figure2}b, we plot $\overline{\omega_\mathrm{2D}}$ and $\overline{\omega_\mathrm{G}}$ for different strain levels in the correlation plot of 2D and G frequency (sample U1). The data points follow a linear trend. This shows that we systematically varied the strain at constant doping. The same linear trend was observed for two other graphene flakes, where strain was induced along the armchair direction (samples U2 and U3 in Fig.\ \ref{fig:Figure2}b; for spectra see Supplementary Fig.\ S2). From a linear fit, we obtain the slope for hydrostatic or biaxial strain $\omega_\mathrm{2D}^\mathrm{h} /\omega_\mathrm{G}^\mathrm{h} = 2.21\pm0.05$. The value is in excellent agreement with theoretical calculations of Mohr et al.\ \cite{Mohr2010} and measured values on graphene blisters reported by Metten et al.\ \cite{Berciaud2014} and Lee et al.\ \cite{Cheong2012}. It also agrees with the experimentally determined slope of $2.45 \pm 0.37$ measured by Zabel et al.\ \cite{Zabel2012} due to its large margin of error.

The peak position for pristine graphene (yellow star) lies outside the measured 2D versus G frequency line in Fig.\ \ref{fig:Figure2}b. This is due to p-doping of the graphene flakes by the PMMA substrate \cite{Pirkle2011, Gammelgaard2014, JiWon2013}. $\omega_\mathrm{2D}/\omega_\mathrm{G}$ for p doping in the absence of strain is shown as a red line in Fig.\ \ref{fig:Figure2}b. We used the expected G frequency shift under p-type doping to estimate the doping of the graphene flakes as $\sim6\times10^{12}$ cm$^{-2}$ (Ref. \citenum{Das2008}, see also Ref.\ \citenum{Froehlicher2015}). 

From the frequency shift and splitting of the G line we calculate the hydrostatic and shear strain giving rise to the spectra in Fig.\ \ref{fig:Figure2}a. We use a Grueneisen parameter of $\gamma_\mathrm{G}=1.8$ \cite{Zabel2012, Ding2010, Mohiuddin2009, Berciaud2014} and shear deformation potential of $\beta_\mathrm{G} = 0.99$ \cite{Mohiuddin2009}. The shear strain increases linearly with hydrostatic strain, which is expected for uniaxial stress (Fig.\ \ref{fig:Figure2}b inset). Hydrostatic strain and shear strain under uniaxial tension are connected by the Poisson ratio $\nu$ as $\varepsilon_\mathrm{s} =  [(1+\nu)/(1-\nu)] \varepsilon_\mathrm{h}$. The Poisson ratio $\nu = 0.33$ is used for the case of ideal stress-strain transfer from the PMMA beam to graphene \cite{Mohiuddin2009, Corro2015}. The expected slope agrees well with the data points extracted for sample U1 and U2 (see dots and squares in the inset of Fig.\ \ref{fig:Figure2}b). For sample U3, the shear strain at a given hydrostatic strain is larger than expected. In this experiment the PMMA beam was deflected several times before the measurement was taken, which led to a non-ideal load resulting in a different strain configuration. Extracting hydrostatic and shear strain components is useful for determining the strain configuration and testing the stress-strain transfer from a substrate to graphene. For samples U1 and U2 we observed good agreement of the uniaxial strain calculated from the beam deflection and the uniaxial strain calculated from the measured hydrostatic strain (Supplementary Fig.\ S3).

\section{Complex nanoscale strain configurations}
\begin{figure*}
\centerline{\includegraphics{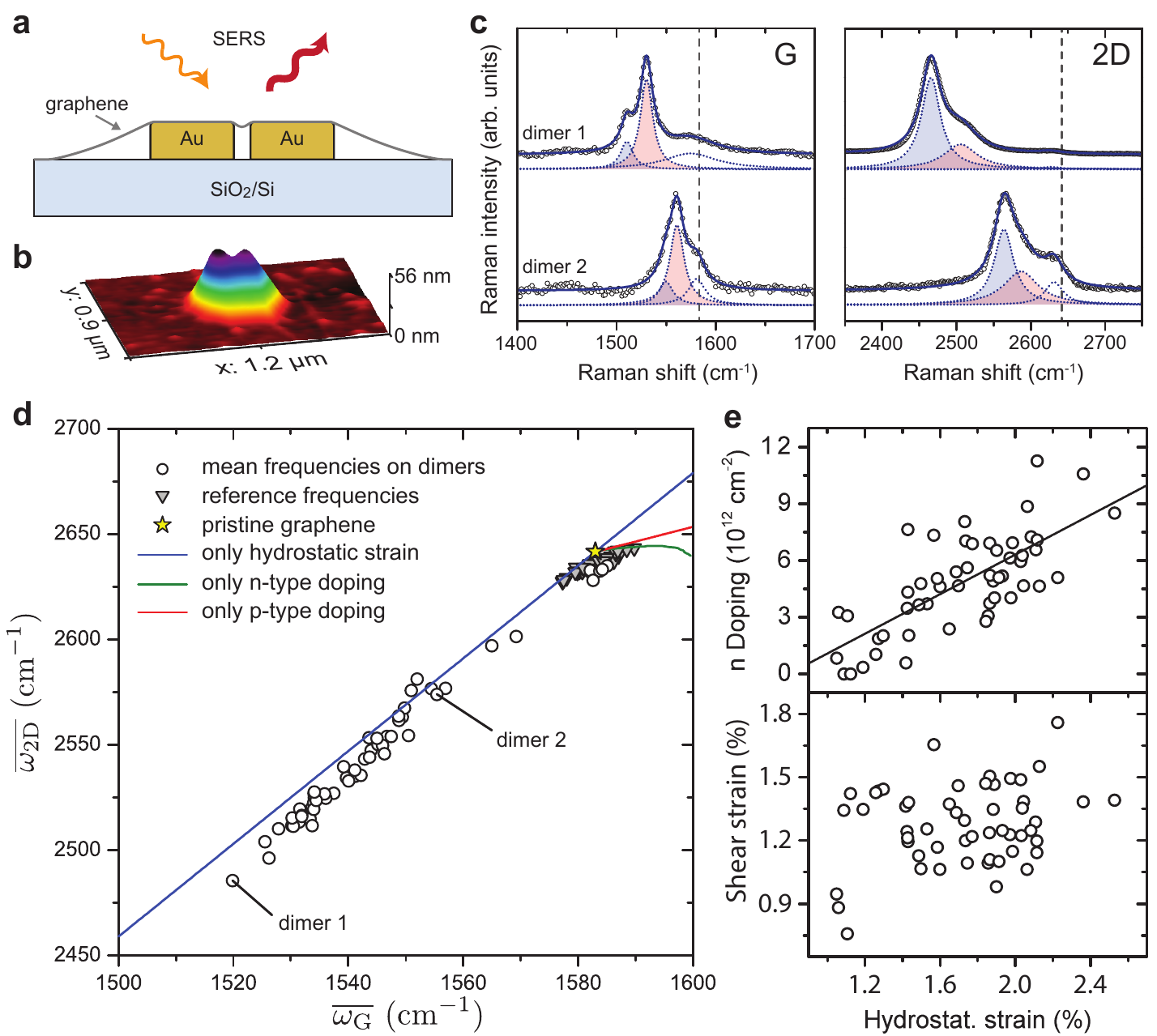}}
\caption{Raman characterization of graphene on gold nanodimers. (a) Geometry of plasmonic nanostructure and substrate. Arrows indicate surface-enhanced Raman scattering (SERS) using graphene as Raman scatterer. (b) Representative 3-dimensional AFM profile of the graphene-covered nanodimer. (c) Exemplary Raman spectra of graphene covering two Au nanodimers with strong (top) and weak (bottom) strain. G$^-$ and 2D$^-$ peaks are colored blue, G$^+$ and 2D$^+$ peaks are colored red. The spectral positions for unstrained graphene are indicated by grey dashed lines (sample D1, laser wavelength 638 nm, linear light pol.\ along dimer axis). (d) Correlation plot of 2D and G mean frequencies (circles). Spectral positions are indicated by triangles for reference measurements on SiO$_2$/Si. The yellow star shows the frequencies for unstrained and undoped graphene (deduced from experimental data in Ref.\ \citenum{Bruna2014}, 633 nm laser excitation, assuming a 2D mode dispersion of 100 cm$^{-1}$/eV). Expected peak positions are shown by the blue line for hydrostatic strain (no doping), the red line for p-type doping (Ref.\ \citenum{Lee2012}, data from Refs.\ \citenum{Das2008,Das2009}; no strain) and the green line for n-type doping\cite{Das2009} (no strain). (e) Extracted magnitude of doping, shear strain and hydrostatic strain induced in graphene by the nanodimers. A linear fit is superimposed on the data in the upper plot.}
\label{fig:Figure3}
\end{figure*}

So far, we presented one application of our methodology for strain evaluation. Using circular light polarization turned out to be useful for unraveling the local strain and doping levels, especially for weak strain when no peak splitting is detectable. As a second example, we analyze complex and unknown strain configurations on the nanoscale. Exfoliated flakes of graphene were suspended on multiple lithographically-fabricated pairs of gold nanodiscs with diameters of $\sim 100$ nm, heights of $40 - 70$ nm and gaps of $20-30$ nm (Fig.\ \ref{fig:Figure3}a) \cite{Heeg2013, Heeg2013RRL}. 
After transfer on top of the nanodimers, the substrate adhesion pulled the graphene into the gap between the two gold nanoparticles. This induced strong local strain with completely unknown configurations (see AFM topography, Fig.\ \ref{fig:Figure3}b). The Raman spectrum from the graphene bridging the dimer gap is strongly enhanced by surface-enhanced Raman scattering \cite{Heeg2013, Heeg2013RRL, Heeg2014}. The plasmon of the gold nanodimer induces strong electromagnetic near fields that increase the local Raman cross section by three to four orders of magnitude \cite{Heeg2013}. The near field enhancement provides nanoscale spatial resolution; it also fixes the polarization of the field to be parallel to the nanodimer axis.

Figure \ref{fig:Figure3}c shows two representative Raman spectra with different levels of intrinsic strain. They belong to the same graphene flake but were recorded on different nanodimers. Both G and 2D modes are split into three components of different intensity. The two peaks with the lowest Raman shift arise from the strained graphene in the interparticle gap \cite{Heeg2013}. The third peak is only slightly shifted from the peak position expected for pristine graphene (grey dashed lines in Fig.\ \ref{fig:Figure3}c). It arises from the unstrained graphene around the nanodimer that contributes to all Raman spectra without plasmonic enhancement. In the following, we focus on the two peaks that are shifted to lower wavenumbers and thereby analyze the local strain in the dimer cavity. We identify these as the G$^-$ and G$^+$ or 2D$^-$ and 2D$^+$ components. The spectra show clear signatures of hydrostatic strain (peak shift) and shear strain (peak splitting).

We now turn to a statistical analysis of the strain and doping locally induced in different positions of a graphene flake by various nanodimers. The dimers were individually addressed with a confocal Raman microscope. We recorded the Raman spectra of 60 graphene-covered gold nanodimers, all of which showed surface-enhanced Raman scattering. We calculated the mean frequencies $\overline{\omega_\mathrm{G}}$ and $\overline{\omega_\mathrm{2D}}$ from fits similar to Fig.\ \ref{fig:Figure3}c for each nanodimer. The contributions of doping and hydrostatic strain were separated with a correlation plot of the mean 2D and G frequencies shown in Fig.\ \ref{fig:Figure3}d. The peak positions are strongly shifted compared to the reference measurements 1 $\mu$m away from each nanodimer (grey triangles) and of pristine graphene (yellow star). To determine the local hydrostatic strain, we needed to analyse the contribution of local charge doping first. We assumed an n-type doping in the graphene bridging the dimer cavity, as previously observed for graphene on gold nanoparticles \cite{Wu2012, Huh2011, Zasada2012, Fang2012}. The $\omega_\mathrm{2D}/\omega_\mathrm{G}$ positions for n-type doping are plotted as a green line in Fig.\ \ref{fig:Figure3}c \cite{Das2009}. The local n-type doping was calculated from the expected G frequency shifts \cite{Das2009}. It reached levels of up to $10^{13}$ cm$^{-2}$, see Fig.\ \ref{fig:Figure3}e. This is more than one order of magnitude higher than the maximum doping levels of 6$\times 10^{11}$ cm$^{-2}$ for a comparable plasmonic system reported in Ref.\ \citenum{Fang2012}. This is expected because we probe the local doping in graphene in the vicinity of the gold nanoparticles, whereas Fang.\ et al.\ \cite{Fang2012} probed the charge doping of the entire graphene sheet by transport measurements. Our approach complements the macroscopic measurements with an opportunity to probe local doping levels in an area of $\sim 20\times20$ nm$^2$. Interestingly, the local charge doping increased statistically with the hydrostatic strain (Fig.\ \ref{fig:Figure3}e, calculated using a Grueneisen parameter of $\gamma_\mathrm{G} = 1.8$ \cite{Zabel2012, Ding2010, Mohiuddin2009, Berciaud2014}). We expect that with increasing local strain, graphene is pulled deeper into the dimer cavity. This leads to a stronger interaction between graphene and the gold nanostructure and explains the increase in doping with strain. 

The nanodimers induced strong local strain in the graphene lattice with a hydrostatic component ranging from 1\% to 2.6\% (Fig.\ \ref{fig:Figure3}e). The shear component was calculated from the G splitting using a shear deformation potential of $\beta_\mathrm{G} = 0.99$ \cite{Mohiuddin2009}. With a magnitude from 0.75\% to 1.8\% it was weaker than the hydrostatic component (Fig.\ \ref{fig:Figure3}e). This differs from uniaxial strain, where the shear component was twice as large as the hydrostatic component (Fig.\ \ref{fig:Figure2}b inset) and biaxial strain where the shear component vanishes. Shear strain and hydrostatic strain are largely uncorrelated for the nanodimers (Fig.\ \ref{fig:Figure3}e), from which we conclude that the nanodimers induce different strain configurations in the suspended graphene. Our methodology for strain evaluation is even applicable for this complicated case because it does not require any assumption of the strain configuration as an input parameter. 

\begin{figure}
\centerline{\includegraphics{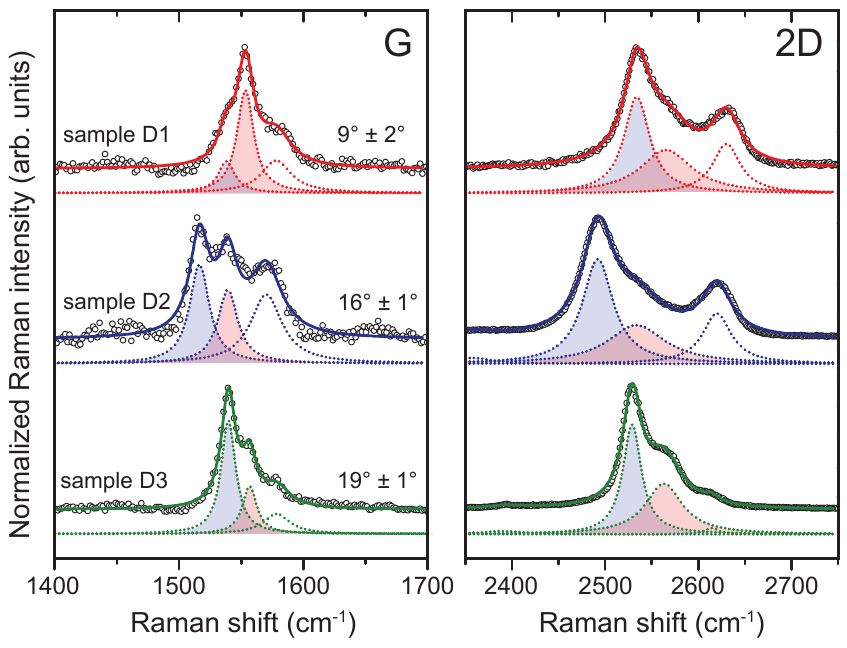}}
\caption{Strain evaluation for several graphene flakes covering gold nanodimers (samples D1-D3, laser wavelength 638 nm, linear light pol.\ along dimer axes). Exemplary Raman spectra are shown for three samples. Fits with three Lorentzian peaks are superimposed on the experimental data and offset for clarity. G$^-$ and 2D$^-$ components are colored blue and G$^+$ and 2D$^+$ components red. The angles between the zigzag direction in the graphene lattice and the dimer axes were calculated from $I_{\mathrm{G}^-}/I_{\mathrm{G}^+}$ and are given next to the spectra. }
\label{fig:Figure4}
\end{figure}

Finally, we demonstrate our analysis of local strain and doping for several different graphene flakes covering arrays of gold nanodimers. Representative Raman spectra of three graphene flakes (samples D1-D3) are shown in Fig.\ \ref{fig:Figure4}a. While the intensity ratio of the 2D$^-$ and 2D$^+$ components is similar in all spectra, the G$^-$/G$^+$-intensity ratio varies from one flake to the other. This is explained by the different lattice orientation of the graphene flakes with respect to the direction of the local shear-strain component (see Methods section) \cite{Huang2009, Mohiuddin2009}. Based on symmetry arguments as discussed in detail in Ref.\ \citenum{Heeg2013} we assume that the local shear strain is directed along and perpendicular to the dimer axis. We determined the angle between the zigzag direction of graphene and the dimer axis for 86 nanodimers covered by five different graphene flakes (Supplementary Fig.\ S4a). The estimated angles range from $(9\pm2)^\circ$ for sample D1 to $(19\pm1)^\circ$ for D3. We also repeated the correlation analysis of the 2D and G mean frequencies for all graphene flakes (Supplementary Fig.\ S4b). The peak positions nicely follow the same trend as for sample D1 in Fig.\ \ref{fig:Figure3}d. We found statistically the same strain and doping levels in all graphene flakes. 

\section{Conclusion}
In conclusion, we proposed a method for analyzing arbitrary strain configurations and simultaneous doping in graphene using Raman spectroscopy. First, the shift due to pure hydrostatic strain is determined through explicit or implicit (circular light) averaging of the G and 2D mode components. Second, the peak shifts induced by hydrostatic strain and doping are separated by a correlation analysis of the G and 2D mean frequencies. This offers the possibility to calculate the local shear strain, hydrostatic strain and doping without any assumption on the underlying strain configuration. We demonstrated this strain analysis for graphene on two different substrates. First, graphene was subjected to uniaxial stress by deflection of a PMMA beam. We demonstrate that circular corotating light polarization is ideal for obtaining the mean frequencies of the G and 2D mode under strain. This approach is even applicable for low strain levels when no peak splitting is visible. As a second example, graphene was suspended on pairs of gold nanoparticles, which induced strong local strain of completely unknown configuration. Surface-enhanced Raman scattering allowed extracting the local strain and doping levels with nanoscale resolution. The two examples serve as model cases for the strain analysis of arbitrary strain configurations in graphene. Our methodology also carries over to other 2D materials, such as MoS$_2$ \cite{Sun2014, Liu2016}.

\section{Methods}
\subsection{Sample fabrication and characterization}
For the measurements under uniaxial strain, monolayer graphene flakes were obtained by micromechanical exfoliation and deposited on a flexible polymethylmethacrylat (PMMA) beam. Prior to graphene transfer, the PMMA substrate was spin coated with SU8 photoresist (SU8 2000.5, MicroChem). The samples were soft-cured at 80 $^\circ$C for 30 min and exposed to UV radiation (366 nm, 30 sec). Graphene was transferred onto the samples with the scotch-tape method. Finally, the samples were covered with PMMA (1\% in anisole) to improve the strain transfer efficiency. Uniaxial stress was induced by bending the PMMA beam with a jig (see e.g.\ Refs.\ \citenum{Tsoukleri2009, Frank2010}). We assume that the doping does not change when inducing strain in graphene by deflection of the PMMA substrate, as we observed no systematic linewidth change of the G peak components with increasing strain. To avoid heating and accompanying structural change of the graphene-PMMA interface, we used laser powers below 150 $\mu$W for Raman characterization (532 nm laser excitation).

The graphene-covered gold nanodimers were fabricated using the same procedure as described in Ref.\ \citenum{Heeg2013}. In short, gold nanodisk dimers were produced with electron-beam lithography, followed by metallization (5 nm Cr or Ti and 40 to 80 nm Au) and lift-off in an ultrasonic bath. Micromechanically cleaved large flakes of single-layer graphene were then transferred on top of the plasmonic nanostructures using a dry transfer method. The spacing between the nanodimers beneath the graphene flakes was much larger than the laser-spot size. This enabled us to record Raman spectra of graphene interacting with an individual nanodimer. 

Raman spectra were recorded with a Horiba XploRA single-grating confocal Raman spectrometer, equipped with a 1200 groves per mm grating, which leads to a spectral resolution of 2-3 cm$^{-1}$. The spectrometer was calibrated with the Raman response of diamond. 532 nm laser excitation was used for measurements on graphene under uniaxial stress ($2\times120$ sec acquisition time, 130 $\mu$W laser power, spot size of 340 nm). Reference spectra of the PMMA substrate were recorded with the same parameters next to the graphene flakes and subtracted from the Raman spectra of graphene. The graphene-covered gold nanodimers were characterized with 638 nm laser excitation ($2\times 120$ sec acquisition time, 280 $\mu$W laser power, spot size of 605 nm); at this wavelength large plasmonic enhancement is expected \cite{Heeg2013}. The laser was focused on each plasmonic nanodimer using a piezo stage and steps of 100 nm in $x$-, $y$- and $z$ direction until a maximum intensity of the 2D mode was achieved. Linear light polarization was chosen along the dimer axis. Reference spectra were recorded 1 $\mu$m away from each nanodimer. 

\subsection{Polarization dependent measurements}
The Raman intensities of the G components can be calculated with the macroscopic theory of Raman scattering as $I_{\mathrm{G}^\pm} \propto \vert \mathbf{e}_\mathrm{i} R_{\mathrm{G}^\pm} \mathbf{e}_\mathrm{s}\vert^2$. $\mathbf{e}_\mathrm{i}$ is the polarization of the incoming light and $\mathbf{e}_\mathrm{s}$ of the Raman-scattered light.  $R_{\mathrm{G}^-}$ is the Raman tensor for the G$^-$ peak and $R_{\mathrm{G}^+}$ for the G$^+$ peak \cite{Huang2009}. For uniaxial stress applied at an angle $\vartheta$ with respect to the crystallographic zigzag direction in graphene, the intensities are calculated with the Raman tensors given in Ref.\ \citenum{Huang2009} as
\begin{equation}
\label{eq:GmodeIntLinLightPol}
\begin{split}
I_{\mathrm{G}^+} &\propto d^2 \cos{\left( \varphi + \psi + 3 \vartheta \right)}^2, \\ I_{\mathrm{G}^-} &\propto d^2 \sin{\left( \varphi + \psi + 3 \vartheta \right)}^2,
\end{split}
\end{equation}
where $d$ is the Raman tensor component of the G mode for unstrained graphene. $\varphi$ is the angle of the incoming light and $\psi$ the angle of the Raman scattered light to the strain axis. From Eq.\ (\ref{eq:GmodeIntLinLightPol}) it is apparent that $I_{\mathrm{G}^-}$ and $I_{\mathrm{G}^+}$ can be very different; for $\varphi + \psi + 3\vartheta = n \cdot \pi/2$ ($n$ integer) one of the components vanishes. The situation is different for circular light polarization with polarization vectors \cite{Reich2001}
\begin{equation}
\mathbf{e}_\circlearrowright = \frac{1}{\sqrt{2}} \left( \begin{array}{c}
1 \\ 
- \mathrm{i}
\end{array}  \right), \quad
 \mathbf{e}_\circlearrowleft = \frac{1}{\sqrt{2}} \left( \begin{array}{c}
1 \\ 
\mathrm{i}
\end{array}  \right).
\end{equation}
For circular corotating polarization (i.e.\ $\circlearrowright\circlearrowright$ or $\circlearrowleft\circlearrowleft$ as combinations for $\mathbf{e}_\mathrm{i}$ and $\mathbf{e}_\mathrm{s}$) we obtain $I_{\mathrm{G}^-} \propto d^2$ and $I_{\mathrm{G}^+} \propto d^2$; the G mode splits into components of equal intensity. When using circular contrarotating polarization (i.e.\ $\circlearrowright\circlearrowleft$ or $\circlearrowleft\circlearrowright$ ), both G mode components vanish.

The polarization dependence of the 2D mode is more complex because it is a second-order Raman process. The intensities of the 2D$^-$ and 2D$^+$ components are determined by double-resonance processes and the simple theoretical treatment that was used for the G mode cannot be applied \cite{Huang2010, Yoon2011, Narula2012}. If incoming- and outgoing light are linearly polarized along the strain axis ($\updownarrow\updownarrow$), $I_{\mathrm{2D}^-}>I_{\mathrm{2D}^+}$. The opposite case $I_{\mathrm{2D}^-}< I_{\mathrm{2D}^+}$ is observed for light polarizations perpendicular to the strain axis $(\leftrightarrow\leftrightarrow)$. In general there is a complicated functional dependence of $I_{\mathrm{2D}^-}$ and $I_{\mathrm{2D}^+}$ on the polarization directions of incoming and outgoing light \cite{Yoon2011}. A reliable fit of the 2D peak components is only possible for a large peak splitting induced by shear strain. For circular light polarization, the overall peak intensity is determined by the selection rules of the Raman process. The 2D mode has symmetry $A_{1g}\oplus E_{2g}$ \cite{Jorio2017}. Under circular corotating polarization processes with $A_{1g}$ symmetry vanish whereas for circular contrarotating polarization processes with $E_{2g}$ symmetry are zero \cite{Reich2001}. This explains the non-vanishing 2D mode for both, circular corotating and contrarotating light polarization. 

For Raman measurements with circular light polarization, we used an optical setup as schematically depicted in Fig.\ \ref{fig:Figure1}a. The laser beam with initially linear light polarization was directed through a quarter-wave plate ($\lambda/4$) to induce circular light polarization. The light was focused onto the sample with an optical microscope which was also used to collect the backscattered light. After the beam splitter (BS) and notch filter (NF), a second quarter-wave plate switched back to linear light polarization. With an analyzer, we selected either the circular-corotating or the circular-contrarotating polarization in the Raman experiment. In front of the spectrometer entrance we placed a half-wave plate ($\lambda/2$) to ensure that the Raman-scattered light entered the spectrometer always with the same light polarization because the spectrometer response depends on polarization. To realize co- and contrarotating configurations, it is also possible to use a single $\lambda/4$ plate between beam splitter and microscope objective if the setup permits \cite{Reich2001}. Combinations of linear light polarization were realized by using $\lambda/2$ plates instead of $\lambda/4$ plates.
\\

\begin{acknowledgements}
The authors thank V.\ Oddone and E.\ Grelich for help on construction of a jig to induce uniaxial strain in graphene. N.S.M.\ thanks Deutsche Telekom Stiftung for financial support. O.F.\ acknowledges the support of Czech Science Foundation (GACR 17-18702S), M.K.\ and M.P.A.\ the support of MSMT ERC-CZ project (LL 1301). S.H. and A.V. acknowledge funding from the Engineering and Physical Sciences Research council grant EP/K016946/1.
\end{acknowledgements}

\bibliography{./References/EvaluatingNonStandardStrain.bib}

\end{document}

\title{Supporting Information: Evaluating arbitrary strain configurations and doping in graphene with Raman spectroscopy}

\author{Niclas S. Mueller}
\email{niclasm@physik.fu-berlin.de}
\affiliation{Freie Universit\"at Berlin, Department of Physics, Arnimallee 14, D-14195 Berlin, Germany}
\author{Sebastian Heeg}
\affiliation{School of Materials, The University of Manchester, Manchester M13 9PL, UK}
\affiliation{Photonics Laboratory, ETH Z\"{u}rich, 8093 Z\"{u}rich, Switzerland}
\author{Miriam Pe\~{n}a Alvarez}
\affiliation{J. Heyrovsk\'{y} Institute of Physical Chemistry, Academy of Sciences of the Czech Republic, Dolej\v{s}kova 3, CZ-18223 Prague 8, Czech Republic}
\author{Patryk Kusch}
\affiliation{Freie Universit\"at Berlin, Department of Physics, Arnimallee 14, D-14195 Berlin, Germany}
\author{S\"oren Wasserroth}
\affiliation{Freie Universit\"at Berlin, Department of Physics, Arnimallee 14, D-14195 Berlin, Germany}
\author{Nick Clark}
\affiliation{School of Materials, The University of Manchester, Manchester M13 9PL, UK}
\author{Fred Schedin}
\affiliation{Centre for Mesoscience and Nanotechnology, The University of Manchester, Manchester M13 9PL, UK}
\author{John Parthenios}
\affiliation{Institute of Chemical Engineering Sciences, Foundation of Research and Technology - Hellas, Patras 26504, Greece}
\author{Konstantinos Papagelis}
\affiliation{Institute of Chemical Engineering Sciences, Foundation of Research and Technology - Hellas, Patras 26504, Greece}
\affiliation{Department of Physics, University of Patras, Patras 26504, Greece}
\author{Costas Galiotis}
\affiliation{Institute of Chemical Engineering Sciences, Foundation of Research and Technology - Hellas, Patras 26504, Greece}
\affiliation{Department of Chemical Engineering, University of Patras, Patras 26504, Greece}
\author{Martin Kalb\'{a}\v{c}}
\affiliation{J. Heyrovsk\'{y} Institute of Physical Chemistry, Academy of Sciences of the Czech Republic, Dolej\v{s}kova 3, CZ-18223 Prague 8, Czech Republic}
\author{Aravind Vijayaraghavan}
\affiliation{School of Materials, The University of Manchester, Manchester M13 9PL, UK}
\affiliation{National Graphene Center, The University of Manchester, Manchester M13 9PL, UK}
\author{Uwe Huebner}
\affiliation{Leibnitz Institute of Photonic Technology, 07745 Jena, Germany}
\author{Roman Gorbachev}
\affiliation{School of Physics \& Astronomy, The University of Manchester, Manchester M13 9PL, UK}
\author{Otakar Frank}
\affiliation{J. Heyrovsk\'{y} Institute of Physical Chemistry, Academy of Sciences of the Czech Republic, Dolej\v{s}kova 3, CZ-18223 Prague 8, Czech Republic}
\author{Stephanie Reich}
\email{reich@physik.fu-berlin.de}
\affiliation{Freie Universit\"at Berlin, Department of Physics, Arnimallee 14, D-14195 Berlin, Germany}

\date{\today}

\maketitle

\begin{figure}
\centerline{\includegraphics{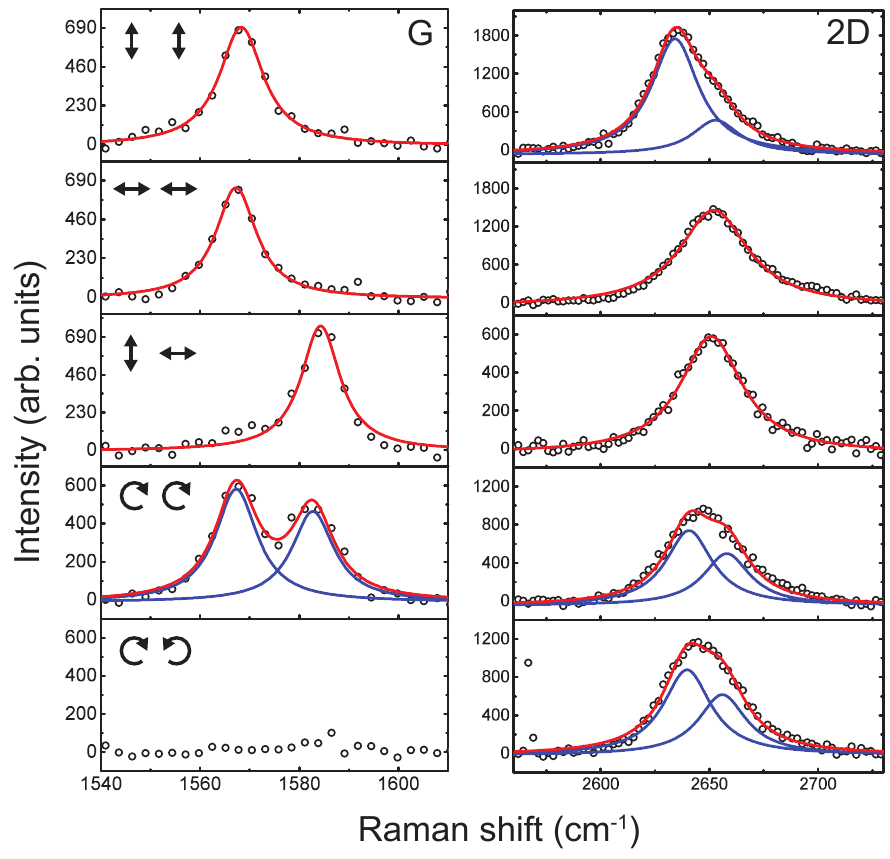}}
\caption{Polarization dependence of the Raman spectrum of graphene with induced uniaxial stress along the armchair direction (sample U2, laser wavelength 532 nm). Fits with one or two Lorentzian peaks are superimposed on the experimental data. Reliable fitting of the 2D mode is complicated by the weak splitting. Polarizations of incoming- and outgoing light are indicated by arrows ($\updownarrow$ - linear along strain, $\leftrightarrow$ - linear perpendicular to strain, $\circlearrowright\circlearrowright$ - circular corotating, $\circlearrowright\circlearrowleft$ - circular contrarotating).}
\end{figure}

\begin{figure}
\centerline{\includegraphics{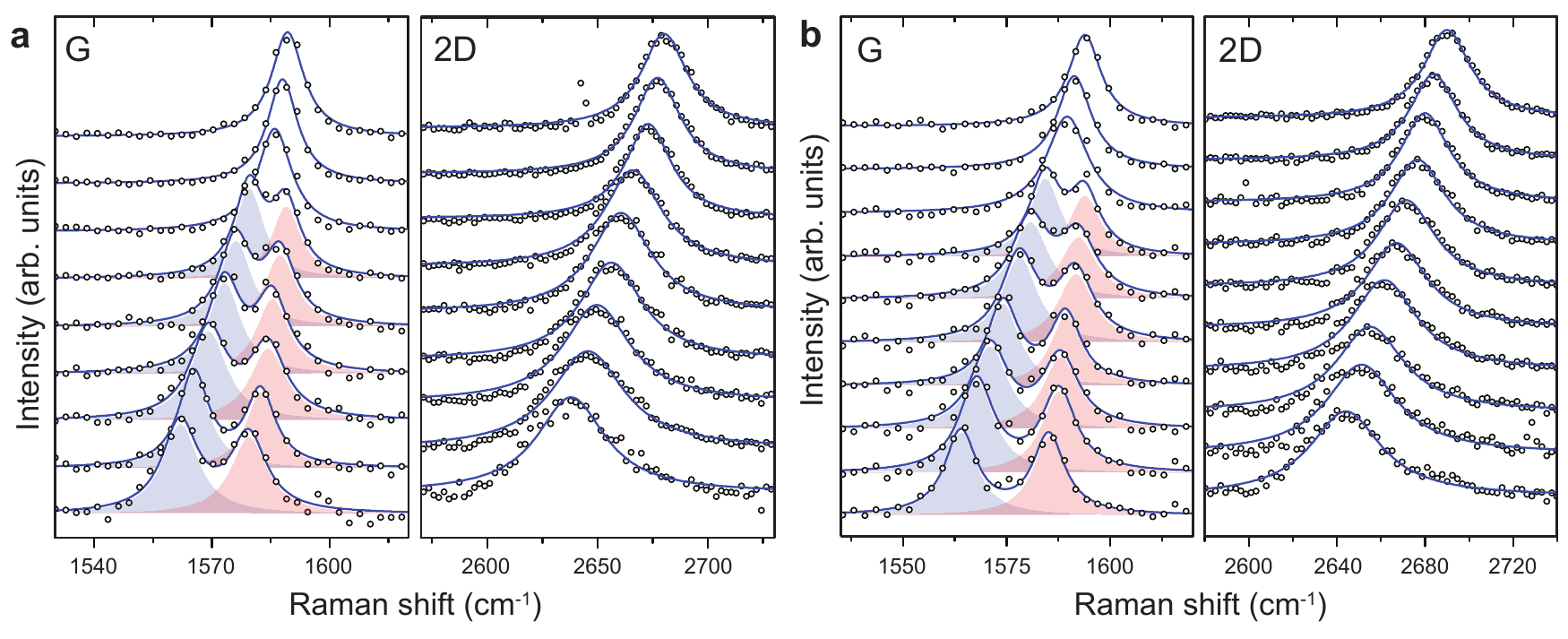}}
\caption{Raman spectra of uniaxially-strained graphene using circular corotating light polarization, for (a) sample U2 and (b) sample U3. The strain was incrementally increased from top to bottom. Fits with two Lorentzian peaks are superimposed on the data. The G$^-$ and 2D$^-$ peaks are colored blue and the G$^+$ and 2D$^+$ peaks are colored red. A single Lorentzian peak fit was used if no peak splitting was visible. The fits were used to obtain the mean frequencies $\overline{\omega_\mathrm{G}}$ and $\overline{\omega_\mathrm{2D}}$ in Fig.\ 3b.}
\end{figure}

\begin{figure}
\centerline{\includegraphics{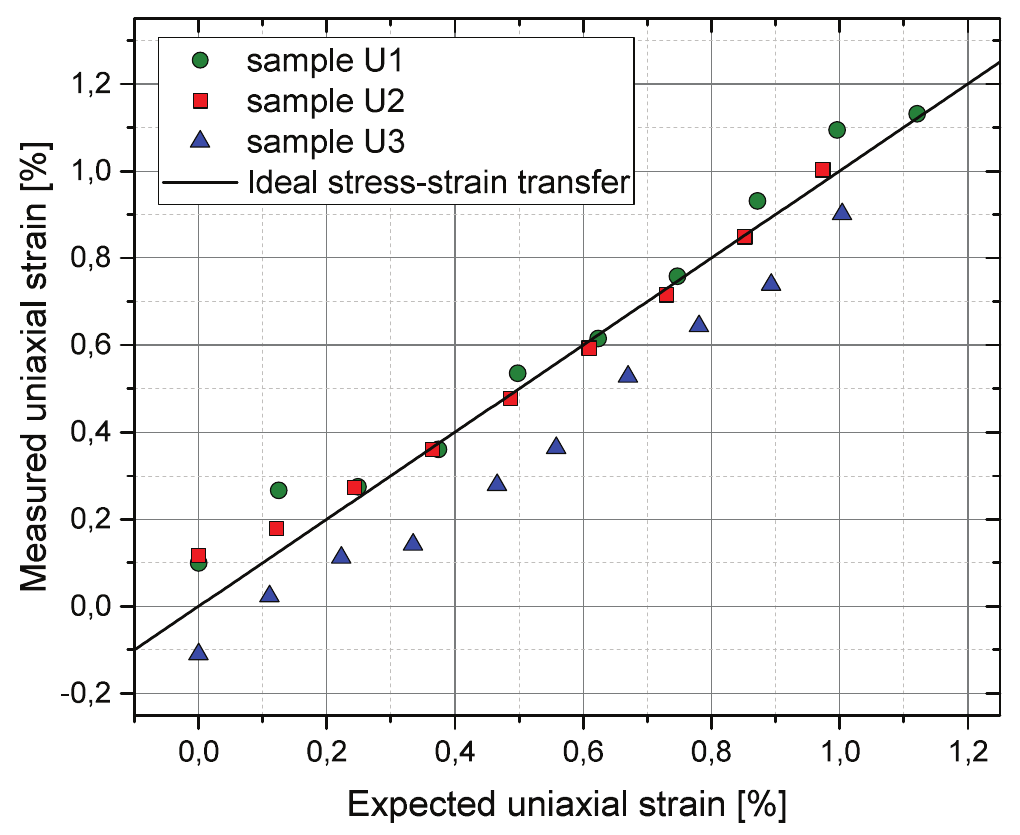}}
\caption{Measured uniaxial strain plotted vs expected uniaxial strain. The measured uniaxial strain $\varepsilon_\mathrm{u}$ was obtained from the measured hydrostatic strain $\varepsilon_\mathrm{h}$ as $\varepsilon_\mathrm{u} = \varepsilon_\mathrm{h}/(1-\nu)$ with a Poisson ratio of $\nu = 0.33$. The expected uniaxial strain was calculated from the beam deflection with the method described in Tsoukleri et al.\ Small 5, 21 (2009). }
\end{figure}

\begin{figure}
\centerline{\includegraphics{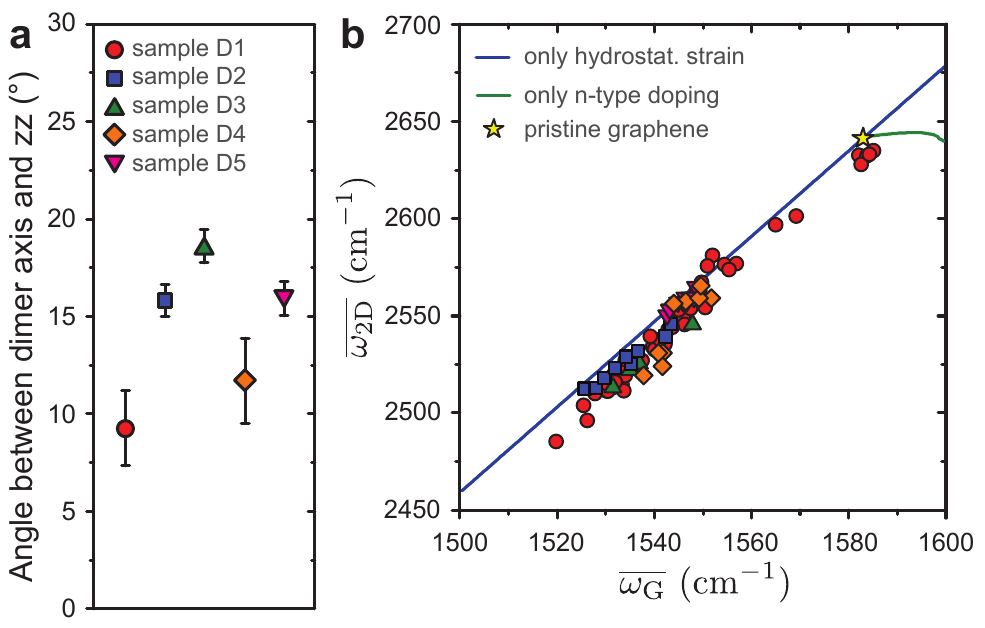}}
\caption{(a) Angle between dimer axes and zigzag direction (zz) in the graphene lattice for five different graphene flakes covering the gold nanodimers (samples D1-D5). (b) Correlation plot with the mean frequencies of G and 2D mode to separate the peak shifts from doping and strain.}
\end{figure}